# A Resource-efficient FIR Filter Design Based on an RAG Improved Algorithm

Mengwei Hu
National Demonstration Center for Experimental Physics and Education
School of Physics and Technology, Wuhan University
Wuhan, China
humengv@whu.edu.cn

Xianyang Jiang*
National Demonstration Center for Experimental Physics and Education
School of Physics and Technology, Wuhan University
Wuhan, China
jiang@whu.edu.cn

Zhengxiong Li
National Demonstration Center for Experimental Physics and Education
Hongyi Honor College of Wuhan University
Wuhan, China
li_zhengxiong@whu.edu.cn

*Abstract*—In modern digital filter chip design, efficient resource utilization is a hot topic. Due to linear phase characteristics of FIR filters, a pulsed fully parallel structure can be applied to attack the problem. In order to further reduce hardware resource consumption especially caused by multiplication function, an improved RAG algorithm is proposed. Filters with different orders and for different algorithms are compared, and the experimental results show that the improved RAG algorithm is excellent in terms of logic resource utilization, resource allocation, running speed, and power consumption under different application scenarios. The proposed algorithm invokes a better circuit structure for FIR filter, it gives full play to resource allocation strategy and reduces logic resource consumption. The proposed circuit is faster and more stable, and suitable for a variety of complex application scenarios.

## I. Introduction

FIR filters have a wide range of applications in communication, audio processing, image processing, and other fields. With blowout type increasement of portable devices and Internet of Things, there is an increasing demand for low power consumption and small size in this field. FIR filter design also evolves in this tide to meet requirements of embedded systems. Efficient FIR filter design methods have been continuously explored to reduce computational cost and product cost, which may involve new optimization algorithms, approximation techniques, and resource allocation strategies.

FIR filters have an important linear phase property, which allows us to exploit the symmetry of coefficients to build both serial and parallel structures. Compared to IIR filters, FIR filters have many significant advantages such as bounded input and output stability, phase linearity, and low coefficient sensitivity, which makes it more suitable under many applications [1]. However, FIR filters involve a large number of arithmetic operations, which limits their processing speed [2]. In order to overcome this limitation, a fully parallel structure can be exploited which allows a single filtering operation to perform multiple multiplications simultaneously to improve performance. Compared to fully parallel FIR filters, adoption of improved Reduced Adder Graph (RAG) algorithm [3] can significantly reduce hardware consumption by exploiting redundancy between coefficients [4]. Our work focuses on how to efficiently implement FIR filters with fixed coefficients. In the proposed design, fully parallel structure and RAG algorithm characteristics have been utilized to effectively reduce hardware cost. Meanwhile, a better resource allocation strategy is taken to further improve FIR filter implementation.

## II. Pulsed fully parallel FIR filters

FIR filters differ from IIR filters in that their impulse response can be expressed in terms of a finite number of sampled values and can be described by a difference equation (1), where N is the number of filters tap coefficients and x(n) is the input time series [5].

$$y(n) = \sum_{k=0}^{N-1} h(k)x(n-k) \qquad (1)$$

In order to implement an efficient specific circuit for FIR, the "pulsation" structure, originally proposed by H.T. Kung, represents a parallel pipelined approach for high-speed signal processing and data processing. This architecture is known for a number of advantages such as modularity, regularity, local linking, and high degree of pipelining. In pulsation architecture, Process Element (PE) constitutes a multiprocessor system and these PEs work together in a synchronized manner so that this architecture offers significant performance in processing large-scale data. The hardware structure of a pulsation FIR filter is shown in Fig. 1, and N PEs are required to accomplish one such operation.

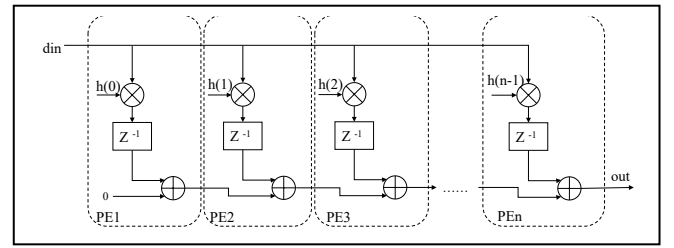

Fig. 1 Hardware structure of pulsation FIR filter

As for an FIR filter with symmetric coefficients, its linear phase property can be further utilized to reduce the number of PEs by pre-addition. For example, a hardware structure of even-symmetric filter is shown in Fig. 2, and it is clear that the number of PEs is reduced to N/2 for the same N tap coefficients.

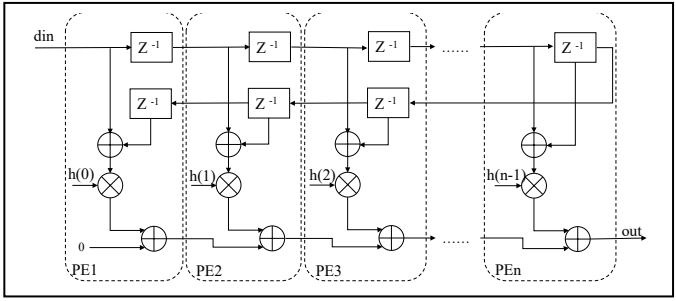

Fig. 2 Structure of pulsation FIR filter with even symmetry of coefficients

III. FILTER IMPLEMENTATION OF RAG IMPROVEMENT ALGORITHM

*A. RAG Improved Algorithm*

Pulsation filter designs are usually based on PEs that have identical coefficients. Since such coefficients are constant and shift operations can be easily implemented in hardware, multiplication can be replaced using shift, add, and subtract operations to reduce resource consumption of multiplier. The concept of using an addition tree in a multiplier was first introduced by Bull, who advocated the implementation of multiplication by constructing a structural graph consisting of simple addition and shift operations, A.G. Dempster and M.D. Macleod proposed RAG algorithm [6], the core idea of this algorithm is to use an equivalent structure to convert all coefficients to bases, and at the same time, introduce subtraction structure to make all intermediate values positive, which greatly simplifies the structure of addition tree. The redundancy relationship between coefficients is also utilized to reduce logical depth in order to cut down the total resource consumption [7,8].

In addition, the RAG algorithm introduces a concept called adder depth "cost". Despite obvious advantages of RAG algorithm, its disadvantage lies in the need to calculate cost value for each coefficient. For smaller coefficients, cost value can be obtained directly by looking up the table, but when coefficients are larger, cost value calculation becomes more difficult [11].

In order to address shortcomings of the RAG algorithm in circuit design, an improved algorithm is proposed, which combines the advantages of pulsation structure fully parallel FIR filter and the RAG algorithm structure FIR filter.

Let "coeff" be all filter coefficients to be realized, "coeff-r" is the set of smaller coefficients, "coeff-s" is the set of larger coefficients. "cost-n" (n=1,2,3,4) is the set of coefficients with different adder depths, "cost-o" is the set of coefficients with other adder depths, and the improved algorithm [9-12] are as follows:

- Take the absolute values of all coefficients and store the results in "coeff" set;
- Remove duplicate coefficients and coefficients with value $2^n$, and the number of remaining coefficients is denoted as N;
- The smaller coefficients are deposited into set "coeff-r", and the number of coefficients deposited is N/2 or (N-1)/2;
- Deposit the remaining larger coefficients into set "coeff-s";
- Divide the even numbers in the "coeff-r" set by $2^n$ to obtain the base;
- Look up the table to get the depth of adder corresponding to each base number, store these coefficients in "cost-n" set, and store the coefficients that cannot be categorized by the table in "cost-o" set;
- Realize coefficients in "cost-1" set;
- Check the sum/difference of coefficients in all realized cost sets, realize the coefficients in higher cost sets by the sum/difference of coefficients and the realized coefficients, and finally realize the coefficients in "cost-o" set;
- Realize the coefficients in "coeff-s" set according to the hardware structure of pulsation FIR filter with symmetric coefficients.

*B. Implementation Example*

Taking a 64th order filter as an example, Fs=250KHz and Fc=20KHz, the filter coefficients after quantization and rounding [13] are shown in Table 1, and due to the symmetry of the coefficients of the FIR filters, only coefficients with 0≤n≤31 need to be discussed here.

TABLE I.   FILTER COEFFICIENTS, H(N) = H(63-N), 32≤N≤63

| h(0)=219 | h(1)=137 | h(2)=162 | h(3)=174 |
|---|---|---|---|
| h(4)=168 | h(5)=137 | h(6)=79 | h(7)=-9 |
| h(8)=-127 | h(9)=-269 | h(10)=-428 | h(11)=-592 |
| h(12)=-747 | h(13)=-875 | h(14)=-957 | h(15)=-972 |
| h(16)=-903 | h(17)=-733 | h(18)=-450 | h(19)=-49 |
| h(20)=470 | h(21)=1100 | h(22)=1825 | h(23)=2622 |
| h(24)=3462 | h(25)=4311 | h(26)=5134 | h(27)=5891 |
| h(28)=6548 | h(29)=7072 | h(30)=7437 | h(31)=7624 |

Firstly, all the coefficients in the above table are taken as absolute values, and duplicates and numbers divisible by $2^n$ are removed, and then the coefficients are divided into the two parts, one part of the coefficients is smaller, which is easy to optimize using the RAG improvement algorithm, and the coefficients are stored into the "coeff-r" set = [9, 49, 79, 127, 137, 162, 168, 174, 219, 269, 428, 450, 470, 592, 733], and one part is larger, and the corresponding coefficients can be stored according to the pulsation structure and the symmetry of the coefficients. 162, 168, 174, 219, 269, 428, 450, 470, 592, 733], parts of the coefficients are larger, according to the pulsation structure and the symmetry of the coefficients, the corresponding input signals can be pre-added or subtracted, and then multiplied by the filter coefficients, which are deposited in the "coeff-s" collection. The coefficients are stored in the set of "coeff-s" = [747, 875, 957, 972, 903, 1100, 1825, 2622, 3462, 4311, 5134, 5891, 6548, 7072, 7437, 7624].

Check the adder depth table to categorize the coefficients and store the coefficients in the corresponding sets:

"cost-1" set = [9, 127];

"cost-2" set = [49, 79, 137, 162, 168];

"cost-3" set = [174, 219];

"cost-o" set = [269, 428, 450, 470, 592, 733].

First realize the coefficients of cost-1:

$$\begin{cases} x9 = xin << 3 + xin, \\ x127 = xin << 7 - xin, \end{cases}$$

Continue to realize the cost-2 factor:

$$\begin{cases} x49 = x9 << 2 + x9 + xin << 2, \\ x79 = x127 - x49 + xin, \\ x137 = x127 + x9 + xin, \\ x162 = x137 + x9 + xin << 4, \\ x168 = x162 + xin << 2 + xin << 1, \end{cases}$$

The cost-3 factor is then realized:

$$\begin{cases} x174 = x127 + x49 - xin << 1, \\ x219 = x168 + x49 - xin << 1, \end{cases}$$

The flexible use of multiplexing makes it possible to use only two layers of adder depth for the coefficients of cost-3 as well, and then gradually implement the other coefficients at the end. The final result of the RAG improvement algorithm design is as follows:

$$\begin{cases} x9 = xin << 3 + xin, \\ x127 = xin << 7 - xin, \end{cases}$$

$$\begin{cases} x49 = x9 << 2 + x9 + xin << 2, \\ x79 = x127 - x49 + xin, \\ x137 = x127 + x9 + xin, \\ x162 = x137 + x9 + xin << 4, \\ x168 = x162 + xin << 2 + xin << 1, \end{cases}$$

$$\begin{cases} x174 = x127 + x49 - xin << 1, \\ x219 = x168 + x49 - xin << 1, \end{cases}$$

$$\begin{cases} x450 = x428 + x9 << 1 + xin << 2, \\ x470 = x450 + x9 << 1 + xin << 1, \\ x592 = x450 + x269 - x127, \\ x733 = x592 + x137 + xin << 2, \end{cases}$$

Comparing to the pre-optimization design, which uses a total of 28 adders, performs 7 shift operations and keeps the adder depth at 2 and below, saves more than half of the total number of adders and a large number of shift operations, while also reducing the adder depth, compared to the unimproved algorithm.

## IV. COMPARISON OF HARDWARE SYNTHESIS RESULTS

The consumption of FPGA hardware resources can be measured by FPGA LUT resources, FF register resources, and DSP resources, and the hardware performance can be measured by power consumption and device junction temperature [9,14]. In our design, we adopt a Virtex-7 series xc7vx485tffg1157-1 FPGA, and implement and compare pulsed fully parallel structure, traditional RAG algorithmic structure, and RAG improved algorithmic structure according to these measuements. The realization results are shown in Table 2 and Figure 3.

TABLE II. COMPARISON OF 64TH ORDER FILTER HARDWARE

| Resource performance indicators | different algorithmic structures | | |
|---|---|---|---|
| | *Pulsed Fully Parallel* | *RAG algorithm* | *RAG Improved Algorithm* |
| LUT | 574 | 4956 | 934 |
| FF | 1286 | 528 | 904 |
| DSP | 4 | 0 | 2 |
| Power(W) | 32.8 | 234.7 | 38.6 |
| Tem(°C) | 70.8 | 125.0 | 79 |

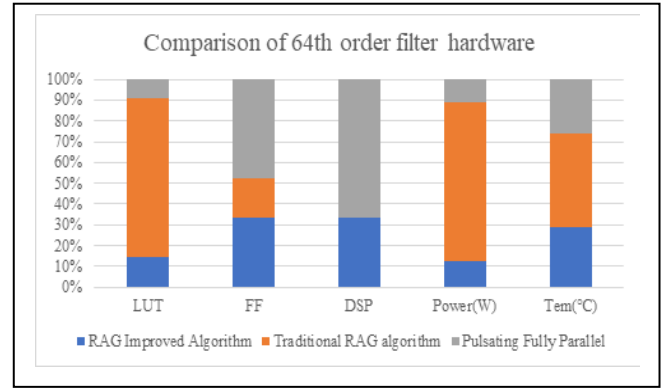

Fig. 3 Comprehensive comparison of 64th order filter hardware

From the results, we can see that the pulsation structure fully parallel filter uses the most DSP and FF and the least LUT resources, while the conventional RAG algorithmic structure filter does not use DSP and consumes the least FF resources, but it is clear that the RAG algorithmic structure uses too much LUT resources compared to the other two structures to the extent that the power and junction temperatures are too high to be used for practical applications. The RAG improved algorithmic structure filter reduces the DSP usage by half and the FF resources by about 29.7% compared to the pulsed fully parallel structure, while the LUT resources are reduced by about 81.2% compared to the RAG algorithmic structure while achieving low power consumption and junction temperature.

Table 3 and 4 show the hardware synthesis comparison of the 8th and 32nd order filters, respectively, to validate the hardware synthesis results for different order filters.

TABLE III. COMPARISON OF 8TH ORDER FILTER HARDWARE

| Resource performance indicators | different algorithmic structures | | |
|---|---|---|---|
| | *Pulsed Fully Parallel* | *RAG algorithm* | *RAG Improved Algorithm* |
| LUT | 141 | 212 | 185 |
| FF | 203 | 120 | 222 |
| DSP | 4 | 0 | 2 |
| Power(W) | 41.762 | 33.673 | 36.75 |
| Tem(°C) | 83.4 | 72.1 | 76.4 |

It can be seen that the traditional RAG algorithm structure has the best integrated performance when the filter order is 8, the RAG improved algorithm structure has excellent performance, and the pulsed fully parallel structure has the worst integrated performance, and there is little difference in the integrated performance of the three except for the DSP resource consumption.

TABLE IV. COMPARISON OF 32ND ORDER FILTER HARDWARE

| Resource performance indicators | different algorithmic structures | | |
|---|---|---|---|
| | *Pulsed Fully Parallel* | *RAG algorithm* | *RAG Improved Algorithm* |
| LUT | 358 | 695 | 555 |
| FF | 679 | 287 | 538 |
| DSP | 4 | 0 | 2 |
| Power(W) | 21.34 | 24.52 | 19.75 |
| Tem(°C) | 54.8 | 59.3 | 52.6 |

While the pulsed fully parallel structure consumes the most FF and DSP resources and the RAG algorithm structure consumes the most LUT resources when the filter order is 32, the LUT, FF, and DSP resource consumption of the RAG improved algorithm structure filter are all in the middle between the pulsed fully parallel structure and the traditional RAG algorithm filter, which are more balanced in logic distribution, and the power consumption and junction temperature are the lowest among the three [15].

Comprehensive comparison concluded that the pulsed fully parallel structure is more suitable for higher order filters, the traditional RAG algorithm structure is more suitable for low order filters, and regardless of the order number, the RAG improved algorithm structure filters have excellent performance, effectively take advantage of the resource allocation strategy, and achieve low power consumption, low junction temperature, and optimal resource consumption for 32-order and 64-order filters. Specifically, the 64th order filter reduces DSP usage and balances logic resource consumption, and stabilizes power consumption and junction temperature as well. The proposed design is very suitable for applications in the case of many coefficients and high complexity.

## V. CONCLUSION

FIR filters exist in more and more application scenarios, and practical requirements are becoming more and more individualized, so novel design solutions must be continuously explored to meet these requirements under various scenarios. The algorithmic structure described provides a new design scheme that is suitable for most application scenarios and is well suited when the number of filter orders is large or the coefficients are large. FIR filter design for improved RAG algorithm gives full play to resource allocation strategy by using shift and add operations instead of a direct multiplier structure. Combining with pulsation fully parallel structure, both operation speed and resource utilization efficiency are enhanced. Comparative simulation experiments demonstrate that the improved RAG algorithm and filter structure save a large amount of logic resources, meet low-power requirements, and decrease contradiction between speed and resource-consumption as well.

ACKNOWLEDGMENT

This work was supported in part by National Science Foundation of China under Grant 61072135, 81971702, the Fundamental Research Funds for the Central Universities under Grant 2042017gf0075, 2042019gf0072, and Natural Science Foundation of Hubei Province under Grant 2017CFB721.